\documentclass[eqsecnum,pra]{revtex4}
\usepackage{graphicx}
\usepackage{graphics,amssymb,bm}

\def\ket#1{|\,#1\,\rangle}

\def\braket#1#2{\langle\, #1\,|\,#2\,\rangle}

\def\ol#1{\overline{#1}}

\def\opone{\leavevmode\hbox{\small1\kern-3.8pt\normalsize1}}

\newcommand{\opx}{\hat{\sigma}_x}
\newcommand{\opy}{\hat{\sigma}_y}
\newcommand{\opz}{\hat{\sigma}_z}
\newcommand{\identity}{\hat I}

\newcommand{\beq}{\begin{equation}}
\newcommand{\eeq}{\end{equation}}
\newcommand{\beqa}{\begin{eqnarray}}
\newcommand{\eeqa}{\end{eqnarray}}

\begin{document}
\title{A new method of construction of all sets of mutually unbiased bases for two-qubit systems}

\author{Iulia Ghiu\footnote{iulia.ghiu@g.unibuc.ro}}

\affiliation{Centre for Advanced Quantum Physics, Department of Physics,
University of Bucharest, PO Box MG-11, R-077125,
Bucharest-M\u{a}gurele, Romania}

\begin{abstract}
Mutually unbiased bases are an important tool in many applications of quantum information theory.
We present a new algorithm for finding the mutually unbiased bases for two-qubit systems. We derive a system of four equations in the Galois field GF(4) and show that the solutions of this system are sufficient for obtaining the most general set of mutually unbiased bases. Further, our algorithm is applied to an example and we show that there are three possible solutions of the system of four equations, each solution leading to a different set of mutually unbiased bases.
\end{abstract}

\maketitle

\section{Introduction}
The concept of mutually unbiased bases (MUBs) was initially proposed by Julian Schwinger \cite{Schwinger}.
These states are maximally incompatible, which means that using a basis to obtain optimal outcomes, leads to maximally random results when the other bases are considered in the measurements. If we denote two elements of different orthonormal bases by $\{\ket{\psi_j}\}$ and $\{\ket{\phi_k}\}$, then the two bases are called mutually unbiased if
$$|\braket{\psi_j}{\phi_k}|^2=\frac{1}{d}.$$
where $d$ is the dimension of the Hilbert space \cite{Wootters}. The MUBs constitute now a basic ingredient in many applications of quantum information processing: quantum tomography, quantum key distribution required in cryptography \cite{cripto}, discrete Wigner function \cite{Wootters-doi}, quantum teleportation \cite{telep}, or quantum error correction codes \cite{err}.

If the measurements used for the state reconstruction are built with the help of the mutually unbiased bases, then this scheme is called MUB tomography \cite{Klimov-R}.
The simplest example of MUBs can be given for qubits. If the system is prepared in the state $\ket{\uparrow }$ (or $\ket{\downarrow }$), then a measurement of the spin along the $x-$ or $y-$axis generates the outcomes {\it up} or {\it down} with the same probability, which is equal to 1/2. The eigenvectors of the Pauli operators $\opx $, $\opy $, and $\opz $  form a set of MUBs for qubits.

It is known that the maximal number of mutually unbiased bases can be at most $d+1$ \cite{Ivanovic}. This maximal value can be reached in the case when the dimension of the space is a prime ($d = p$) or a power of prime ($d = p^n$) \cite{Calder}.

It was proved in the case when $d$ is a power of a prime number that there is an approach for finding the MUBs, namely one has to generate families of unitary operators, whose eigenvectors solve the problem of MUBs. In other words, one needs to determine $d + 1$ classes of $d - 1$ commuting operators. These special operators are called mutually unbiased operators \cite{Band}.

The paper is organized as follows: In Section II we present the standard set of MUBs for qubits, where three bases are obtained from tensor product of $\hat \sigma_j $ and $\identity $. Further, some basic notations and definitions used in the Galois fields are recalled in Section III. The main result of the paper is given in Section IV, where we find a system of four equations in the Galois field GF(4). The solution of these equations leads to the construction of the most general set of MUBs for two-qubit systems, not only the standard one. Finally, we make some concluding remarks in Section V.

\section{Preliminaries}

In this paper we analyze the two-qubit systems.
An example of 5 sets of 3 commuting operators for 2 qubits is shown in Table \ref{standard} \cite{Bjork}.

\begin{table}
\begin{center}
\begin{tabular}{|c|c|c|c|} \hline
1. & $\opz \opz $ & $\identity \opz$ &
$\opz \identity$ \\
\hline
2. & $\opx \opx $ & $\identity  \opx $
&
$\opx \identity$ \\
\hline
3. & $\opy \opy$ & $ \identity \opy $ &
$\opy  \identity $ \\
\hline
4. & $\opx \opy$ & $\opz \opx $ &
$\opy \opz $ \\
\hline
5. & $\opy \opx $ & $\opz  \opy $ &
$ \opx \opz$   \\
\hline
\end{tabular}
\end{center}
\caption{The standard set of MU operators for two qubits.}
\label{standard}
\end{table}

We did not write the tensor multiplication sign in Table \ref{standard}. In the case of two-qubit systems, there is only one allowed structured of the bases, namely (3,2), which means that three bases are separable, while two are entangled.
The three separable bases given by Table \ref{standard} are obtained as the common eigenvectors of tensor product of the operators $\opz $ and $\identity $ (all the three possibilities: see the first row of Table \ref{standard}), tensor product of $\opx $ and $\identity $ (second row), and tensor product of $\opy $ and $\identity $ (third row), respectively. The table of operators with this property are called the standard set of MU operators \cite{Klimov-R}.
The last two rows of Table \ref{standard} generate entangled bases which are called {{\it belle}} and {{\it beau}} \cite{Wootters-ibm}.

\section{Galois fields: notations and definitions}

If $d=p^n$ is the power of a prime number, then the $d+1$ MUBs can be constructed with the help of Galois fields GF($d$).
Wootters \cite{Wootters-ibm} and Gibbons {\it et all.} \cite{Gibbons} proposed a method to associate the MUBs to the so-called discrete phase space.
The discrete phase space of a d-level system is a $d \times d$ lattice, whose coordinates are elements of the finite Galois field GF($d$) \cite{Wootters-ibm}. Further a state is associated to a line in the discrete phase space. The set of parallel lines is called a striation. There are $d+1$ striations. It turns out that the MUBs are determined by the bases associated with each striation.

In the following we will recall some basic definitions used in Galois fields.
The map trace of a field element $x \in $ GF($p^n$) is as follows:
$$\mbox{tr}\; x =x +x^2 +... +x^{p^{n-1}}. $$

A basis $\{ \alpha_i\}$ is called self-dual if $\mbox{tr}  (\alpha_i \alpha_j)  = \delta_{ij}$.

Let us consider two elements $x$, $y\in $GF$(p^n)$ which can be written with the help of two bases $E = \{ e_1,..., e_n\} $ and $F = \{ f_1,..., f_n\} $ as follows:
\beqa
x&=&\sum_{i=1}^nx_{e_i}\, e_i;\nonumber \\
y&=&\sum_{j=1}^ny_{f_j}\, f_j.
\eeqa

We denote by $\opx , \opz $ the generalized Pauli operators \cite{Klimov-Romero}:
\beqa
\opx \ket{n}&=&\ket{n+1}; \nonumber \\
\opz \ket{n}&=&\omega ^n\ket{n}, \nonumber
\eeqa
where $\ket{n}$ is the computational basis and $\omega =\exp (2\pi i/d)$ is the $d$th root of the unity.

 To the translation in the discrete phase space by the element $(x,y)$ we associate an operator $T_{(x,y)}$ called the translation operator \cite{Gibbons}:
\beq
T_{(x,y)}:= \opx ^{x_{e_1}}\, \opz ^{y_{f_1}}\otimes ... \otimes \opx ^{x_{e_n}}\, \opz ^{y_{f_n}}.
\label{coresp}
\eeq

In the case when $d=2^2$, there is only one irreducible polynomial: $P(x)=x^2+x+1$. We denote by $\mu $ the primitive element of GF(4) and we get
$$\mu^2 +\mu +1=0. $$
The elements of GF(4) are:
$$\{ 0, 1, \mu , \mu^2 \}.$$

If we use the selfdual basis $\{ \mu,\mu^2\}$ in GF(4), then the following correspondence given by Eq. (\ref{coresp}) between the set of points in the discrete phase space and the set of MU operators can be established:
\beqa
(\mu,0) & \longleftrightarrow & \opx \identity ,
\nonumber \\
(\mu^2,0) & \longleftrightarrow & \identity \opx , \label{op-puncte}  \\
(0,\mu ) & \longleftrightarrow & \opz \identity  ,
\nonumber \\
(0,\mu^2) & \longleftrightarrow & \identity \opz .\nonumber
\eeqa

\section{The new algorithm of construction of MUBs}

\subsection{The commutation relations}

Two operators commute if their associated points in the phase-space $(a_1,b_1)$ and $(a_2,b_2)$ satisfy \cite{Klimov-Romero}

\begin{equation}
\mbox{tr}(a_1b_2)=\mbox{tr}(a_2b_1).
\label{rel-com-gen}
\end{equation}

The MUBs can be constructed with the help of $d+1$ classes of $d-1$ commuting operators, i.e. MU operators. In the case of two-qubit systems, we need to generate a table with 5 rows, each row representing the set of 3 commuting operators. We denote the operator which belongs to the row $r$ and column $c$, where $r=\ol{1,5}$ and $c=\ol{1,3}$ by $O_{r,c}$ as one can see in the left Table of Eq. (\ref{general}).
The point in the phase space $(a^{(r)}_c,b^{(r)}_c)$ is associated to the operator $O_{r,c}$:
$$ O_{r,c}\longleftrightarrow   (a^{(r)}_c,b^{(r)}_c).$$ The correspondence between the set of MU operators and the set of points in the discrete phase space is shown below:

\beq
\begin{tabular}{|c|c|c|c|} \hline
1.& $O_{1,1} $ & $O_{1,2}$& $O_{1,3}$ \\
\hline
2.& $O_{2,1} $ & $O_{2,2}$& $O_{2,3}$ \\
\hline
3.&$O_{3,1} $ & $O_{3,2}$& $O_{3,3}$ \\
\hline
4.&$O_{4,1} $ & $O_{4,2}$& $O_{4,3}$ \\
\hline
5.&$O_{5,1} $ & $O_{5,2}$& $O_{5,3}$ \\
\hline
\end{tabular}
\hspace{0.5cm} \longleftrightarrow  \hspace{0.5cm}
\begin{tabular}{|c|c|c|c|} \hline
1. & $\left( a_1^{(1)},b_1^{(1)}\right) $ & $\left( a_2^{(1)},b_2^{(1)}\right) $   & $\left( a_3^{(1)},b_3^{(1)}\right) $   \\
\hline
2. & $\left( a_1^{(2)},b_1^{(2)}\right) $ & $\left( a_2^{(2)},b_2^{(2)}\right) $   & $\left( a_3^{(2)},b_3^{(2)}\right) $   \\
\hline
3. & $\left( a_1^{(3)},b_1^{(3)}\right) $ & $\left( a_2^{(3)},b_2^{(3)}\right) $   & $\left( a_3^{(3)},b_3^{(3)}\right) $   \\
\hline
4. & $\left( a_1^{(4)},b_1^{(4)}\right) $ & $\left( a_2^{(4)},b_2^{(4)}\right) $   & $\left( a_3^{(4)},b_3^{(4)}\right) $   \\
\hline
5. & $\left( a_1^{(5)},b_1^{(5)}\right) $ & $\left( a_2^{(5)},b_2^{(5)}\right) $   & $\left( a_3^{(5)},b_3^{(5)}\right) $   \\
\hline
\end{tabular}
\label{general}
\eeq
\\

There are four operators which define the whole Table \ref{standard}, namely the first two operators of the rows 1 and 2: $O_{1,1} $; $O_{1,2}$;  $O_{2,1} $; $O_{2,2}$. The other 11 operators are determine as follows \cite{Bjork}:

$O_{r,c} = O_{r,c-2} O_{r,c-1}$, for $r=1, 2$; $c=3$ and

$O_{r,c} = O_{2,c} O_{1,c+r-3}$  for $r>2$.\\
The indices $r$ and $c$ are written modulo four.

In our new construction, we consider that the right Table of Eq. (\ref{general}) of points in the discrete phase space which correspond to the set of MU operators is unique defined by the following four points:
\beqa
&&(a^{(1)}_1,b^{(1)}_1); \;  (a^{(1)}_2,b^{(1)}_2);\nonumber\\
&&(a^{(2)}_1,b^{(2)}_1); \;  (a^{(2)}_2,b^{(2)}_2).
\label{puncte}
\eeqa

Therefore, we need to determine the eight parameters $a^{(i)}_j$ and $b^{(i)}_j$ given by Eq. (\ref{puncte}) such that all the 15 points which are obtained correspond to MU operators. In other words, we have to use the commutation relation in the discrete phase space given by Eq. (\ref{rel-com-gen}) for the set of 3 points of all the 5 rows. For each row we write 3 commutation relations, which lead to a total number of 15 equations for the whole table of points. These 15 equations are written in Appendix A.

\subsection{The general algorithm}

We proved that only 4 equations of the total number of 15 are independent. These are Eqs. (\ref{ec-1}), (\ref{ec-4}), (\ref{ec-7}), and (\ref{ec-10}) in Appendix A:
\beqa
&&\mbox{tr}\left[ a^{(1)}_1b^{(1)}_2\right]=\mbox{tr}\left[ a^{(1)}_2b^{(1)}_1\right]\nonumber\\
&&\mbox{tr}\left[ a^{(2)}_1b^{(2)}_2\right]=\mbox{tr}\left[ a^{(2)}_2b^{(2)}_1\right] \label{ec-sistem} \\
&&\mbox{tr}[ a^{(1)}_1b^{(2)}_2+ a^{(2)}_1b^{(1)}_2]=\mbox{tr}[ a^{(1)}_2b^{(2)}_1+a^{(2)}_2b^{(1)}_1] \nonumber\\
&&\mbox{tr}[ a^{(2)}_1b^{(1)}_1+ a^{(2)}_1b^{(1)}_2+a^{(1)}_2b^{(2)}_2]=\mbox{tr}[ a^{(2)}_2b^{(1)}_2+a^{(1)}_1b^{(2)}_1+a^{(1)}_2b^{(2)}_1] .\nonumber
\eeqa

The general algorithm of the construction:
\begin{itemize}
\item The parameters $b^{(1)}_1, b^{(1)}_2, b^{(2)}_1, a^{(1)}_1, a^{(2)}_1$ are arbitrary chosen.
\item The three variables $a^{(1)}_2$, $a^{(2)}_2$, and $b^{(2)}_2$ are obtained from the system of equations (\ref{ec-sistem}).
\item After obtaining the solution, one has to find the table of points in the discrete phase space analogue to the right table of Eq. (\ref{general}).
\item With the help of the correspondence given by Eq. (\ref{op-puncte}), one generate the table which contain 5 classes of 3 commuting operators.
\item The eigenvectors of each set of 3 commuting operators represent the MUBs.
\end{itemize}

\subsection{Example}
According to the algorithm presented in Sec. IV.B, we have to fix five parameters:
\beqa
(a^{(1)}_1,b^{(1)}_1)&=&(\mu^2,\mu )\nonumber \\
(a^{(2)}_1,b^{(2)}_1)&=&(\mu ,\mu )\nonumber \\
b^{(1)}_2&=&\mu ^2 . \nonumber
\eeqa

The system of Eqs. (\ref{ec-sistem}) becomes:
\beqa
&& \mbox{tr}\left[ \mu \, a^{(1)}_2\right]=1\nonumber \\
&&\mbox{tr}\left[ \mu \, b^{(2)}_2\right]=\mbox{tr}\left[ \mu \, a^{(2)}_2\right] \label{sist-ex} \\
&&\mbox{tr}[ \mu^2b^{(2)}_2+ 1]=\mbox{tr}[ \mu \, a^{(1)}_2+\mu \, a^{(2)}_2]\nonumber \\
&&\mbox{tr}[ \mu^2+1+a^{(1)}_2b^{(2)}_2]=\mbox{tr}[ \mu^2 a^{(2)}_2+1+\mu \, a^{(1)}_2] .\nonumber
\eeqa

There are three possible solutions of the system (\ref{sist-ex}):

Solution A): $a^{(1)}_2=1$, $a^{(2)}_2=\mu ^2$, and $b^{(2)}_2=\mu^2$. This solution leads to the Table \ref{standard}, i.e. the standard set of MU operators.

Solution B): $a^{(1)}_2=1$, $a^{(2)}_2=1$, and $b^{(2)}_2=\mu$. Therefore we obtain the following table of points in the discrete phase space and accordingly the set of MU operators:

\beq
\begin{tabular}{|c|c|l|}
\hline \hline No. & Set of points & MU operators \\
\hline  1. & $(1,\mu^2); (\mu ,1); (\mu^2,\mu ) $ & $\opx \opy ; \; \; \opy \opz ; \; \; \opz \opx  $ \\
\hline  2. & $(1,\mu ); (\mu ,\mu ); (\mu^2,0) $ &  $\opy \opx ; \; \; \opy \identity ; \; \; \identity \opx  $ \\
\hline  3. & $(0,1); (1,1); (1,0) $  & $\opz \opz ; \; \; \opy \opy ; \; \; \opx \opx   $ \\
\hline  4. & $(0,\mu ); (\mu^2 ,1); (\mu^2 ,\mu^2 ) $  &  $\opz \identity ; \; \; \opz \opy ; \; \; \identity \opy   $  \\
\hline  5. & $(0 ,\mu^2 ); (\mu ,\mu^2 ); (\mu ,0) $ & $\identity \opz ; \; \; \opx \opz ; \; \; \opx \identity  $ \\
\hline \hline
\end{tabular}
\label{tabel-solB}
\eeq

Solution C): $a^{(1)}_2=\mu$, $a^{(2)}_2=1$, and $b^{(2)}_2=\mu$. The table of points in the discrete phase space and the set of MU operators are shown below:

\beq
\begin{tabular}{|c|c|l|}
\hline \hline No. & Set of points & MU operators \\
\hline  1. & $(1,1); (\mu ,\mu^2); (\mu^2,\mu ) $ & $\opy \opy ; \; \; \opx \opz ; \; \; \opz \opx  $ \\
\hline  2. & $(1,\mu ); (\mu ,\mu ); (\mu^2,0) $ &  $\opy \opx ; \; \; \opy \identity ; \; \; \identity \opx  $ \\
\hline  3. & $(0,1); (\mu ,1); (\mu^2,1) $  & $\opz \opz ; \; \; \opy \opz ; \; \; \opz \opy   $ \\
\hline  4. & $(0,1); (0 ,\mu ); (0,\mu^2 ) $  &  $\opz \opz ; \; \; \opz \identity ; \; \; \identity \opz   $  \\
\hline  5. & $(1 ,\mu^2 ); (\mu ,0); (\mu^2 ,\mu^2) $ & $\opx \opy ; \; \; \opx \identity ; \; \; \identity \opy  $ \\
\hline \hline
\end{tabular}
\label{tabel-solC}
\eeq

\section{Conclusions}
In this paper we have presented a new algorithm of construction of all sets of MUBs for two-qubit systems. The main result is the system of four equations (\ref{ec-sistem}) in the Galois field GF(4). With the help of the solution of this system, one can obtain different sets of MUBs, not only the well known standard set.

\section*{Acknowledgments}
I wish to thank Gunnar Bj\"{o}rk, Andrei B. Klimov, and Luis L. S\'anchez-Soto for useful discussions. This work was supported by CNCSIS - UEFISCSU, postdoctoral research project PD code 151, no. 150/30.07.2010 for the University of Bucharest.

\appendix
\section{The 15 commutation relations}

\beqa
&&\mbox{tr}\left[ a^{(1)}_1b^{(1)}_2\right]=\mbox{tr}\left[ a^{(1)}_2b^{(1)}_1\right] \label{ec-1}\\
&&\mbox{tr}\left[ a^{(1)}_1\left( b^{(1)}_1+b^{(1)}_2\right) \right]=\mbox{tr}\left[ b^{(1)}_1\left( a^{(1)}_1+a^{(1)}_2\right) \right] \\
&&\mbox{tr}\left[ a^{(1)}_2\left( b^{(1)}_1+b^{(1)}_2\right) \right]=\mbox{tr}\left[ b^{(1)}_2\left( a^{(1)}_1+a^{(1)}_2\right) \right] \\
&&\mbox{tr}\left[ a^{(2)}_1b^{(2)}_2\right]=\mbox{tr}\left[ a^{(2)}_2b^{(2)}_1\right] \label{ec-4}\\
&&\mbox{tr}\left[ a^{(2)}_1\left( b^{(2)}_1+b^{(2)}_2\right) \right]=\mbox{tr}\left[ b^{(2)}_1\left( a^{(2)}_1+a^{(2)}_2\right) \right] \\
&&\mbox{tr}\left[ a^{(2)}_2\left( b^{(2)}_1+b^{(2)}_2\right) \right]=\mbox{tr}\left[ b^{(2)}_2\left( a^{(2)}_1+a^{(2)}_2\right) \right] \\
&&\mbox{tr}\left[ \left( a^{(1)}_1+ a^{(2)}_1\right) \left( b^{(1)}_2+b^{(2)}_2\right) \right]=\mbox{tr} \left[ \left( a^{(1)}_2+ a^{(2)}_2\right) \left( b^{(1)}_1+b^{(2)}_1\right) \right] \label{ec-7}\\
&&\mbox{tr}\left[ \left( a^{(1)}_1+ a^{(2)}_1\right) \left( b^{(1)}_3+b^{(2)}_3\right) \right]=\mbox{tr} \left[ \left( a^{(1)}_3+ a^{(2)}_3\right) \left( b^{(1)}_1+b^{(2)}_1\right) \right]\\
&&\mbox{tr}\left[ \left( a^{(1)}_2+ a^{(2)}_2\right) \left( b^{(1)}_3+b^{(2)}_3\right) \right]=\mbox{tr} \left[ \left( a^{(1)}_3+ a^{(2)}_3\right) \left( b^{(1)}_2+b^{(2)}_2\right) \right]\\
&&\mbox{tr}\left[ \left( a^{(2)}_1+ a^{(1)}_2\right) \left( b^{(2)}_2+b^{(1)}_3\right) \right]=\mbox{tr} \left[ \left( a^{(2)}_2+ a^{(1)}_3\right) \left( b^{(2)}_1+b^{(1)}_2\right) \right] \label{ec-10} \\
&&\mbox{tr}\left[ \left( a^{(2)}_1+ a^{(1)}_2\right) \left( b^{(2)}_3+b^{(1)}_1\right) \right]=\mbox{tr} \left[ \left( a^{(2)}_3+ a^{(1)}_1\right) \left( b^{(2)}_1+b^{(1)}_2\right) \right]\\
&&\mbox{tr}\left[ \left( a^{(2)}_2+ a^{(1)}_3\right) \left( b^{(2)}_3+b^{(1)}_1\right) \right]=\mbox{tr} \left[ \left( a^{(2)}_3+ a^{(1)}_1\right) \left( b^{(2)}_2+b^{(1)}_3\right) \right]\\
&&\mbox{tr}\left[ \left( a^{(2)}_1+ a^{(1)}_3\right) \left( b^{(2)}_2+b^{(1)}_1\right) \right]=\mbox{tr} \left[ \left( a^{(2)}_2+ a^{(1)}_1\right) \left( b^{(2)}_1+b^{(1)}_3\right) \right]\\
&&\mbox{tr}\left[ \left( a^{(2)}_1+ a^{(1)}_3\right) \left( b^{(2)}_3+b^{(1)}_2\right) \right]=\mbox{tr} \left[ \left( a^{(2)}_3+ a^{(1)}_2\right) \left( b^{(2)}_1+b^{(1)}_3\right) \right]\\
&&\mbox{tr}\left[ \left( a^{(2)}_2+ a^{(1)}_1\right) \left( b^{(2)}_3+b^{(1)}_2\right) \right]=\mbox{tr} \left[ \left( a^{(2)}_3+ a^{(1)}_2\right) \left( b^{(1)}_2+b^{(1)}_1\right) \right] .
\eeqa

We proved that these 15 equations are not independent, which means that if we start we the Eqs. (\ref{ec-1}), (\ref{ec-4}), (\ref{ec-7}), and (\ref{ec-10}) we obtain all the other 11 equations. The set of four equations is the minimal one, that generates the other equations.

\end{document}